\documentclass[fleqn,twoside]{article}
\usepackage{amsmath,espcrc2,graphicx}
\title{Nuclear modification of valence-quark distributions
       and its effects \\ on NuTeV $\sin^2 \theta_W$ anomaly}
\author{M. Hirai
     \address{Institute of Particle and Nuclear Studies, KEK,
              1-1, Ooho, Tsukuba, Ibaraki, 305-0801, Japan},
     S. Kumano
     \address[saga]{Department of Physics, Saga University,
              Saga, 840-8502, Japan}
     \thanks{Address after Jan. 1, 2005: 
             KEK, 1-1, Ooho, Tsukuba, 305-0801, Japan},
     and T.-H. Nagai
     \addressmark[saga]
     }
\begin{document}
\begin{abstract}
We investigated a nuclear modification difference between up- and
down-valence quark distributions by analyzing structure function $F_2$ 
and Drell-Yan cross-section ratios. Although nuclear modifications of
the valence-quark distributions themselves are rather well determined,
it is difficult to find their difference from the present data.
We estimated such an effect on the NuTeV $\sin^2 \theta_W$ value
and its uncertainty by the Hessian method. At this stage,
it is not large enough to explain the whole NuTeV anomaly.
However, the modification difference cannot be precisely determined,
so that further studies are needed.
\vspace{1pc}\end{abstract}
\maketitle

\section{Introduction}

Weak-mixing angle $\sin^2 \theta_W$ is one of the fundamental constants
in the standard model. It has been measured experimentally by various 
methods such as atomic parity violation and polarization asymmetry in
electron-positron annihilation. The NuTeV collaboration determined it
by neutrino- and antineutrino-nucleus scattering and found that its value,
$\sin^2 \theta_W = 0.2277 \pm 0.0013 \, \text{(stat)} 
                    \pm 0.0009 \, \text{(syst)}$ in the on-shell scheme,
is significantly larger than other measurements 
($\sin^2 \theta_W= 0.2227 \pm 0.0004$) \cite{nutev}. 
It is called ``NuTeV $\sin^2 \theta_W$ anomaly". 

Because there may be new physics behind this anomaly, careful analyses
are needed for clarifying the situation. In our work, we investigate
a conservative explanation without exotic mechanisms. The NuTeV target
is iron, so that nuclear corrections could be a candidate for
the anomalous result. There are various factors including the effects
of neutron excess, strange-antistrange asymmetry,
isospin violation, and modification of valence-quark
distributions. Among them, the nuclear modification difference
between $u_v$ and $d_v$ is discussed in this paper.
It was first investigated in Ref. \cite{sk} and a detailed analysis
has been done in Ref. \cite{hkn}. Here, we report recent analysis results.

This paper consists of the following.
In section \ref{analysis}, our analysis method and results are explained
for determining the nuclear modification difference.
Its effect on $\sin^2 \theta_W$ is calculated in section \ref{sinth}.
The results are summarized in section \ref{summary}.

\section{Nuclear modification of valence-quark distributions}
\label{analysis}

It is known that nuclear parton distribution functions (PDFs) are modified 
from those of the nucleonic distributions. Such nuclear modifications
have been investigated especially in the structure function $F_2$.
Determination of each parton modification is not straightforward
from $F_2$ and Drell-Yan data. However, gross features of nuclear PDFs
are now determined, for example, in Ref. \cite{npdf04}. Among the nuclear
PDF corrections, it was pointed out that valence-quark modifications
affect the $\sin^2 \theta_W$ determination \cite{sk}.

In order to discuss such a nuclear effect, we define
nuclear modification factors $w_{u_v}$ and $w_{d_v}$ 
for up- and down-valence quark distributions by
\begin{align}
u_v^A (x) & = w_{u_v} (x,A,Z) \, \, \frac{Z u_v (x) + N d_v (x)}{A},
\nonumber \\
d_v^A (x) & = w_{d_v} (x,A,Z) \, \, \frac{Z d_v (x) + N u_v (x)}{A},
\label{eqn:wpart}
\end{align}
where $u_v (x)$ and $d_v (x)$ are the distributions in the nucleon,
$Z$ is the atomic number of a nucleus, and $A$ is the mass number.
For simplicity, the $Q^2$ dependence is abbreviated.
It should be noticed that these relations are used at any $Q^2$
for the present research, although similar equations are defined
only at $Q^2$=1 GeV$^2$ in Ref. \cite{npdf04}. These equations are
motived by the following considerations. If there were no nuclear
correction, the $u_v$ distribution of a nucleus is given
by the simple summation of proton and neutron contributions
$Z u_v^p + N u_v^n$. The isospin symmetry is usually assumed for
the parton distributions, so that it becomes $Z u_v + N d_v$.
It is divided by $A$ ($(Z u_v + N d_v)/A$) because we use nuclear
PDFs per nucleon. Therefore, the function $w_{u_v}$ indicates
nuclear correction to this distribution.

In particular, we are interested in the modification
difference $w_{u_v}-w_{d_v}$ and its effect on the $\sin^2 \theta_W$
determination. Therefore, specific parameters
($a_v'$, $b_v'$, $c_v'$, $d_v'$) are assigned for the difference
at $Q^2$=1 GeV$^2$ \cite{hkn}:
\begin{align}
w_{u_v} (x, & A,Z) -  w_{d_v} (x,A,Z)
= \left( 1 - \frac{1}{A^{1/3}} \right)
\nonumber \\
& \times
\frac{a_v' (A,Z) +b_v' x+c_v' x^2 +d_v' x^3}{(1-x)^{\beta_v}}.
\label{eqn:dwv}
\end{align}
Because of the baryon-number and charge conservations, the parameter
$a_v'$ is fixed, and then the number of parameters is three.
The parameter $\beta_v$ and other parameters in $w_{u_v}+ w_{d_v}$,
$w_{\bar q}$, and $w_g$ are fixed at the values of our recent
analysis \cite{npdf04}.
The mass-number dependence is assumed in the $1 - 1/A^{1/3}$ form
simply by considering nuclear volume and surface contributions
\cite{npdf04}. The $x$ dependence is motivated by the shape of
nuclear modification data of $F_2$. However, we should aware that
appropriate $A$ and $x$ dependence is not known almost at all
for $w_{u_v} - w_{d_v}$.

\begin{figure}[t!]
\begin{center}
     \includegraphics[width=0.40\textwidth]{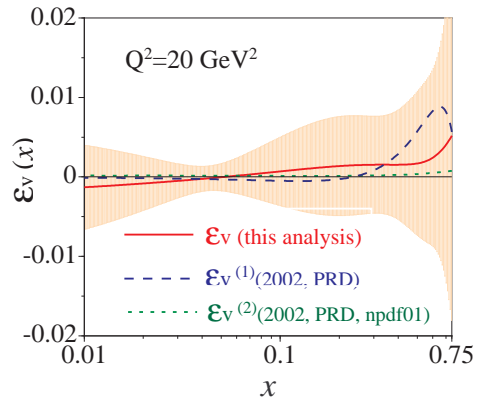}
\end{center}
\vspace{-1.3cm}
\caption{The obtained distribution $\varepsilon_v (x)$ is shown 
         at $Q^2$=20 GeV$^2$ \cite{hkn} with previous results \cite{sk}.
         The details are explained in the text.}
\vspace{-0.4cm}
\label{fig:epsv}
\end{figure}

We determined the parameters by using experimental data for
the nuclear $F_2$ ratios and Drell-Yan cross-section ratios as
investigated in Ref. \cite{npdf04}. The parameters
$b_v'$, $c_v'$, and $d_v'$ are optimized by a $\chi^2$ analysis with the data.
The obtained distribution is shown by the ratio
\begin{equation}
\varepsilon_v (x) = \frac{w_{d_v}(x,A,Z)-w_{u_v}(x,A,Z)}
                             {w_{d_v}(x,A,Z)+w_{u_v}(x,A,Z)}
\, ,
\label{eqn:ev}
\end{equation}
in Fig. \ref{fig:epsv}. Equation (\ref{eqn:ev}) is used at any $Q^2$.
The distribution $\varepsilon_v (x)$ is shown by the solid curve,
and the shaded area indicates the one-$\sigma$ error range. The error
is estimated by the Hessian method by using the determined parameters
and error matrix in the $\chi^2$ analysis. The distribution is shown
at $Q^2$=20 GeV$^2$ which is about the average $Q^2$ value of the NuTeV
data.

The dashed and dotted curves indicate the estimated distributions
in Ref. \cite{sk}. The dashed curve is obtained by calculating
$\varepsilon_v = - (N-Z) (u_v-d_v) (w_v-1)/[A (u_v+d_v)w_v]$,
which is one of the candidates for satisfying the baryon-number
and charge conservations. The function $w_v$ is given by
$w_v=(w_{u_v}+w_{d_v})/2$. The dotted is obtained by Eq. (\ref{eqn:ev})
with the 2001 version of the nuclear PDFs \cite{npdf04}. As obvious
from the figure, three distributions are much different. However,
they are well within the error band, which indicates that these results
are consistent each other. 

\section{Effects on NuTeV $\sin^2 \theta_W$}
\label{sinth}

From the neutrino- and antineutrino-nucleon scattering,
$\sin^2 \theta_W$ could be obtained by using 
the Paschos-Wolfenstein (PW) relation:
\begin{equation}
R^-  = \frac{  \sigma_{NC}^{\nu N}  - \sigma_{NC}^{\bar\nu N} }
              {   \sigma_{CC}^{\nu N}  - \sigma_{CC}^{\bar\nu N} }
        =  \frac{1}{2} - \sin^2 \theta_W 
\, ,
\label{eqn:pw}
\end{equation}
where $\sigma_{CC}^{\nu N}$ and $\sigma_{NC}^{\nu N}$ are
charged-current (CC) and neutral-current (NC) cross sections.
There are various nuclear corrections to the PW relation.
Expanding the expression in terms of the correction factors,
we obtain a modified PW relation for a nucleus \cite{sk}:
\begin{align}
& R_A^-   =  \frac{1}{2} - \sin^2 \theta_W  
\nonumber \\
       &  
- \varepsilon_v (x)  \bigg \{ \bigg ( \frac{1}{2} - \sin^2 \theta_W \bigg ) 
               \frac{1+(1-y)^2}{1-(1-y)^2} - \frac{1}{3} \sin^2 \theta_W 
\bigg \}
\nonumber \\
       &  
+O(\varepsilon_v^2)+O(\varepsilon_n)+O(\varepsilon_s)+O(\varepsilon_c) 
\, .
\label{eqn:apw3}
\end{align}
The first correction term with $\varepsilon_v (x)$ is investigated
in this paper. In addition, there are corrections
associated with the neutron-excess ($\varepsilon_n$), 
strange-antistrange asymmetry ($\varepsilon_s$), and
charm-anticharm asymmetry ($\varepsilon_c$) factors.
The variable $y$ is defined by $y=q^0/E_{\nu}$ 
with the energy transfer $q^0$. 

Because the $\varepsilon_v$ correction term
depends on the variables $x$, $y$, and $Q^2$,
we need to take the NuTeV kinematics into account to estimate
an effect on the $\sin^2 \theta_W$ determination. In particular,
there are few experimental data in the large $x$ region, where
the distribution $\varepsilon_v (x)$ becomes significant as shown in
Fig. \ref{fig:epsv}. Namely, the $\varepsilon_v$ effect on $\sin^2 \theta_W$
could be much smaller than the one expected from Fig. \ref{fig:epsv}.
Fortunately, such kinematical factors are supplied in Ref. \cite{nutev-prd}.
Comparing our PDF definition with the NuTeV one, we find the relations
\begin{align}
\delta u_v^* = u_{vp}^*-d_{vn}^* = 
          - \varepsilon_v \, (w_{u_v}+w_{d_v}) \, x \, u_v ,
\nonumber \\
\delta d_v^* = d_{vp}^*-u_{vn}^* = 
          + \varepsilon_v \, (w_{u_v}+w_{d_v}) \, x \, d_v ,
\label{eqn:pdfdef}
\end{align}
where the asterisk ($*$) indicates the NuTeV distributions. Therefore, 
the nuclear modification difference $\varepsilon_v$ corresponds
to the isospin violation in the NuTeV terminology. Then, the contribution
to the NuTeV $\sin^2 \theta_W$ is given by 
\begin{align}
\Delta (\sin^2 \theta_W) = 
- \int dx \, & \big\{ \, F[\delta u_v^*,x] \, \delta u_v^* (x)
\nonumber \\
             & + F[\delta d_v^*,x] \, \delta d_v^* (x) \, \big\} ,
\label{eqn:del-sinth}
\end{align}
where $F[\delta u_v^*,x]$ and $F[\delta d_v^*,x]$ are the functionals
provided in Fig. 1 of Ref. \cite{nutev-prd}. 
Our sign convention of $\Delta (\sin^2 \theta_W)$ is opposite to
the NuTeV one in Eq. (\ref{eqn:del-sinth}). The distributions
$\delta u_v^* (x)$ and  $\delta d_v^* (x)$ are calculated by
Eq. (\ref{eqn:pdfdef}) with the distribution $\varepsilon_v (x)$, which was 
obtained in the previous section.
Calculating the integral numerically, we obtain \cite{hkn}
\begin{equation}
\Delta (\sin^2 \theta_W) = 0.0004 \pm 0.0015 ,
\label{eqn:effect}
\end{equation}
at $Q^2$=20 GeV$^2$. 

In comparison with the NuTeV deviation $0.0050$,
the correction is an order of magnitude smaller. However, the error becomes
comparable to the deviation. The magnitude of the error depends much
on the analysis condition. Therefore, careful analyses could alter
the values in Eq. (\ref{eqn:effect}). In particular, the nuclear modification
difference $w_{u_v} - w_{d_v}$, which we have investigated in this paper,
is not a uniquely determined quantity at this stage. In order to determine
this difference, we need future experimental efforts.

\section{Summary}
\label{summary}

We have analyzed the experimental data on nuclear structure functions $F_2^A$
and Drell-Yan cross sections for extracting the nuclear modification
difference between $u_v$ and $d_v$ distributions. Using the obtained
distribution, we investigated its effect on the NuTeV $\sin^2 \theta_W$
determination. We found a rather small effect; however, the uncertainty on
the $\sin^2 \theta_W$ is not small in comparison with the NuTeV deviation.
Because such as nuclear modification is difficult to be determined
at this stage, we need further studies to pin down the modification.




\begin{thebibliography}{9}
\bibitem{nutev}   G. P. Zeller {\it et al.},
                     Phys. Rev. Lett. 88 (2002) 091802.
\bibitem{sk}      S. Kumano,  Phys. Rev. D66 (2002) 111301.
\bibitem{hkn}     M. Hirai, S. Kumano, and T.-H. Nagai, 
                     report SAGA-HE-205-04.
\bibitem{npdf04}  M. Hirai, S. Kumano, and T.-H. Nagai,
                     Phys. Rev. C70 (2004) 044905. 
                  See also M. Hirai, S. Kumano, and M. Miyama,
                     Phys. Rev. D64 (2001) 034003.
\bibitem{nutev-prd} G. P. Zeller {\it et al.}, 
                     Phys. Rev. D65 (2002) 111103.
\end{thebibliography}
\end{document}